\documentclass[aps,prd,twocolumn,superscriptaddress,showpacs,floatfix,10pt,nofootinbib]{revtex4}
\usepackage{amssymb}
\usepackage{textcomp}
\usepackage{graphicx}
\usepackage{epsfig}
\usepackage{amsmath}
\usepackage{slashed}
\usepackage{units}
\usepackage{hyperref}
\usepackage{array}
\usepackage{lmodern}
\usepackage{booktabs}

\begin{document}

\title{Bounds on ortho-positronium and $J/\psi$ -- $\Upsilon$ quarkonia invisible decays \\and constraints on hidden braneworlds in a $SO(3,1)$-broken 5D bulk}
\author{Micha\"{e}l Sarrazin}
\email{michael.sarrazin@ac-besancon.fr}
\thanks{Corresponding author}
\affiliation{Institut UTINAM, CNRS/INSU, UMR 6213, Universit\'{e} Bourgogne--Franche-Comt%
\'{e}, 16 route de Gray, F-25030 Besan\c con Cedex, France}
\affiliation{Laboratory of Analysis by Nuclear Reactions, Department of Physics,
University of Namur, 61 rue de Bruxelles, B-5000 Namur, Belgium}
\author{Coraline Stasser}
\affiliation{Laboratory of Analysis by Nuclear Reactions, Department of Physics,
University of Namur, 61 rue de Bruxelles, B-5000 Namur, Belgium}

\begin{abstract}
While our visible Universe could be a 3-brane, some cosmological scenarios
consider that other 3-branes could be hidden in the extra-dimensional bulk.
Matter disappearance toward a hidden brane is mainly discussed for neutron
-- both theoretically and experimentally -- but other particles are poorly
studied. Recent experimental results offer new constraints on positronium or
quarkonium invisible decays. In the present work, we show how a two-brane
Universe allows for such invisible decays. We put this result in the context
of the recent experimental data to constrain the brane energy scale $M_B$
(or effective brane thickness $M_B^{-1}$) and the interbrane distance $d$
for a relevant two-brane Universe in a $SO(3,1)$-broken 5D bulk. Quarkonia
present poor bounds compared to results deduced from previous
passing-through-walls-neutron experiments for which scenarios with $M_B <
2.5 \times 10^{17}$ GeV and $d > 0.5$ fm are excluded. By contrast,
positronium experiments can compete with neutron experiments depending on
the matter content of each brane. To constrain scenarios up to the Planck
scale, positronium experiments in vacuum cavity should be able to reach $%
\text{Br}(\text{o-Ps} \rightarrow \text{invisible}) \approx 10^{-6}$.
\end{abstract}

\maketitle

Many cosmological scenarios consider the existence of hidden braneworlds,
specifically to explain the origin of dark matter and dark energy or as
alternatives to cosmic inflation \cite{r1,r2,r3,r4,r5,r6,r7,cf1,cf2,cf3,BWED}%
. Among the cosmological models of interest, universes involving $SO(3,1)$%
-broken 5D bulks containing at least two braneworlds are under consideration 
\cite{cf1,cf2,cf3}. In the last decade, it has been shown that such
scenarios imply matter exchange between branes, which is a possible way to
test these models \cite{coupl,M4xZ2,pheno,npm,npmth,exp}. In this context,
neutron disappearance (reappearance) toward (from) a hidden brane has been
widely discussed both theoretically \cite{coupl,M4xZ2,pheno} and
experimentally \cite{npm,npmth,exp}. Nevertheless, other kinds of particles
have been somewhat neglected, although with good reason. In the present
paper, we consider recent experimental results about constraints on
positronium \cite{IPD1,Psin} or quarkonium \cite{IQD1} invisible decays in vacuum \footnote{%
In a previous work \cite{otherC}, the constraint on $\text{Br}(\text{o-Ps} \rightarrow \text{invisible})$ is stronger by 3 orders of magnitude when compared to the work here under consideration \cite{IPD1}. However, this constraint \cite{otherC} is obtained from measurements
performed in presence of matter such that o-Ps undergoes very high collision
rates \cite{IPD1,otherC}. Then, decay rate calculation requires extrapolations \cite{IPD1,otherC,cont}, to account
of collisions consequences, leading to large uncertainties \cite{IPD1,otherC}. As a result,
there is a need for experiments with low collision rates -- or in vacuum -- to avoid any extrapolation \cite{IPD1}.
Nevertheless, uncertainties could be restrained thanks to complex GEANT4 simulations
possibly supplemented with a density matrix approach, as done for
neutron-hidden neutron dynamics in nuclear reactors \cite{npm,npmth}, for instance. This
significant further work is far beyond the topic of the present work aiming to demonstrate the relevance of positronium experiments to track hidden
braneworlds. Thus, for now, we chose to consider the conservative but
robust constraint from experiments in vacuum cavity only \cite{IPD1}.}. These works \cite{IPD1,Psin} were themselves motivated by the possibility of invisible decays along extra dimensions \cite{ThPs}, a situation somewhat similar but different from the present one involving hidden braneworlds. Here, showing how a hidden
braneworld allows for such invisible decays, we discuss about constraints on
the distance between our visible brane and a hidden one, and on the brane
energy scale (or brane thickness) in the bulk. Highlighting the significant
efficiency of the neutron to probe the existence of hidden braneworld, we
put theoretical constraints on the expected branching ratio for positronium
and quarkonium invisible decays.

When considering a two-brane Universe, whatever its full high-energy
description (i.e. whatever the number or properties of bulk scalar fields
responsible for particle trapping on branes, the number of extra dimensions
or the bulk metric, etc.), the fermion dynamics on both branes at a sub-GeV
energy scale is the same as the dynamics of fermions in a $M_{4}\times Z_{2}$
space-time in the context of the non-commutative geometry \cite{coupl,M4xZ2}. There are then two copies of the Standard Model,
each sector being localized in each brane. Assuming that each braneworld is
a $M_{B}^{-1}$-thick domain wall -- where $M_{B}$ is the brane energy scale
-- the two sectors are mutually invisible to each other at the zeroth-order
approximation for processes with energies below $M_{B}$. By contrast, matter
fields in separate branes can mix at a first-order approximation mainly
through the Lagrangian $\mathcal{L}_{c}=ig\overline{\psi }_{+}\gamma
^{5}\psi _{-}+ig\overline{\psi }_{-}\gamma ^{5}\psi _{+}$, where $\psi _{\pm
}$ are the Dirac fermionic fields in each braneworld -- denoted $(+)$ and $%
(-)$ \cite{coupl,M4xZ2}. The interbrane coupling $g$ is given by \cite{coupl}%
: 
\begin{equation}
g=(m^{2}/M_{B})\exp (-md),  \label{cg}
\end{equation}
and depends on the distance $d$ between branes, on their effective thickness 
$M_{B}^{-1}$ and on the mass $m$ of the particle of the Standard Model under
consideration, i.e. a constituent quark \cite{coupl,CQM1,CQM2,CQM3,CQM4} or
a lepton. More specifically for bound quarks, $g$ presents a cut-off at $d
\approx 0.5$ fm beyond which it cancels \cite{coupl}.

From $\mathcal{L}_{c}$ one can show that a particle could oscillate between
two states, one localized in our brane, the other localized in the hidden
world. In fact, the oscillation would be driven by the effective magnetic
field $\mathbf{B}_{\bot }\mathbf{=}g\left( \mathbf{A}_{+}-\mathbf{A}%
_{-}\right) $ transverse-like to the branes, where $\mathbf{A}_{\pm }$ are
the magnetic vector potentials in each brane. Specifically, the interaction
Hamiltonian $\mathbf{H}_{c}$ between the Pauli spinors of the visible and
hidden worlds is given by \cite{coupl,M4xZ2,pheno}: 
\begin{equation}
\mathbf{H}_{c}=\hbar \Omega \left( 
\begin{array}{cc}
0 & \mathbf{\varepsilon } \\ 
\mathbf{\varepsilon }^{\dagger } & 0%
\end{array}%
\right) ,  \label{Coupling}
\end{equation}%
where $\mathbf{\varepsilon }=-i\mathbf{\sigma }\cdot \mathbf{B}_{\bot
}/B_{\bot }=-i\mathbf{\sigma }\cdot \mathbf{n}$ is a unitary matrix acting
on the spin with $\mathbf{n} = (\sin\theta \cos\varphi, \sin\theta
\sin\varphi, \cos\theta )$, and $\hbar \Omega =\mu _{n}B_{\bot }$ with $\mu
_{n}$ the magnetic moment of the particle. Here vector potentials $\mathbf{A}%
_{\pm }$ are dominated by the huge ($\gtrsim 10^{9}$ T m) overall
astrophysical magnetic vector potential $\mathbf{A}_{amb}$ related to the
magnetic fields of all the astrophysical objects (planets, stars, galaxies,
etc.) \cite{pheno,npm,vp1,vp2,vp3,vp4,vp5} and such that $\mathbf{A}_{+}-%
\mathbf{A}_{-}\approx\mathbf{A}_{amb}$ \cite{pheno,npm}.\newline

\begin{table*}[t]
\centering
\renewcommand{\arraystretch}{1.5} 
\begin{tabular}{lr|c|c|c}
$p\overline{p}$ & state & Mass & Full width $2\Gamma$ & $\text{Br}(\text{o-}p%
\overline{p}\rightarrow \text{invisible})$ \\ \hline\hline
Positronium & o-Ps & $1.022-\varepsilon $ MeV & $7.211\times 10^{6}$ s$^{-1}$
(th) & $<5.9\times 10^{-4}$ at$\ 90$ \% C.L. \\ 
& p-Ps & $1.022-\varepsilon ^{\prime }$ MeV & $8.033\times 10^{9}$ s$^{-1}$
(th) & - \\ 
&  &  &  &  \\ 
Charmonium & $J/\psi $ & $3096.900$ $\pm $ $0.006$ MeV & $92.9$ $\pm $ $2.8$
keV (exp) & $<7\times 10^{-4}$ at$\ 90$ \% C.L. \\ 
& $\eta _{c}$ & $2983.4$ $\pm $ $0.5$ MeV & $31.8$ $\pm $ $0.8$ MeV (exp) & -
\\ 
&  &  &  &  \\ 
Bottonium & $\Upsilon $ & $9460.30\pm 0.26$ MeV & $54.02\pm 1.25$ keV (exp)
& $<3.0\times 10^{-4}$ at$\ 90$ \% C.L. \\ 
& $\eta _{b}$ & $9399.0\pm 2.3$ MeV & $10_{-4}^{+5}$ MeV (exp) & - \\ \hline
\end{tabular}
\caption{Summary of positronium and quarkonium masses, full widths and
invisible decay branching ratios from theoretical and experimental data \cite{IPD1,IQD1,CQM1,CQM2,CQM3,CQM4,Qk}. $\varepsilon$
and $\varepsilon ^{\prime}$ describe the mass difference between the
ortho-state and the para-state of the positronium, such that $\left| \varepsilon - \varepsilon ^{\prime} \right| = 0.84$ meV.}
\label{t1}
\end{table*}

In the present work we consider bound states $\overline{p}p$ constituted by
a fermion $p$ and its anti-particle $\overline{p}$. These states are
characterized by a finite lifetime and some decay channels. More precisely,
we consider the long-living $1^{3}S_{1}$ state (ortho-state) and the
short-living $1^{1}S_{0}$ state (para-state) of the $p\overline{p}$ bound
pair. Following the standard description, the ortho-state wave function can
be expressed as \cite{Psqu}: 
\begin{eqnarray}
\left\vert \psi _{1}\right\rangle &=&\left\vert p,\uparrow \right\rangle
\otimes \left\vert \overline{p},\uparrow \right\rangle ,  \label{wf1} \\
\left\vert \psi _{-1}\right\rangle &=&\left\vert p,\downarrow \right\rangle
\otimes \left\vert \overline{p},\downarrow \right\rangle ,  \label{wf-1} \\
\left\vert \psi _{0}\right\rangle &=&(1/\sqrt{2})\left( \left\vert
p,\uparrow \right\rangle \otimes \left\vert \overline{p},\downarrow
\right\rangle +\left\vert p,\downarrow \right\rangle \otimes \left\vert 
\overline{p},\uparrow \right\rangle \right) ,  \label{wf0}
\end{eqnarray}%
and similarly for the para-state wave function, we have: 
\begin{equation}
\left\vert \psi _{P}\right\rangle =(1/\sqrt{2})\left( \left\vert p,\uparrow
\right\rangle \otimes \left\vert \overline{p},\downarrow \right\rangle
-\left\vert p,\downarrow \right\rangle \otimes \left\vert \overline{p}%
,\uparrow \right\rangle \right) ,  \label{wfp}
\end{equation}%
where $\left\vert p,\updownarrow \right\rangle $ and $\left\vert \overline{p}%
,\updownarrow \right\rangle $ relate to the wave functions of the fermion
and its anti-particle respectively both taking into account the spin state.

Let us now consider the relevant interbrane coupling, which is simply the
sum of two $\mathbf{H}_{c}$ Hamiltonian operators corresponding to the
contributions of the particle and its anti-particle \cite{Psqu}:

\begin{equation}
\mathbf{W}=-i\hbar \Omega \left( 
\begin{array}{cc}
0 & \left( \mathbf{\sigma }_{p}\mathbf{-\sigma }_{\overline{p}}\right) 
\mathbf{\cdot n} \\ 
-\left( \mathbf{\sigma }_{p}-\mathbf{\sigma }_{\overline{p}}\right) \mathbf{%
\cdot n} & 0%
\end{array}%
\right) ,  \label{Wff}
\end{equation}%
where the minus sign ($-\mathbf{\sigma }_{\overline{p}}$) arises from the
opposite magnetic moment of these particles. Considering the non-diagonal
terms of the coupling term $\mathbf{W}$, and the $\overline{p}p$ wave
functions, we get: 
\begin{equation}
\left( \mathbf{\sigma }_{p}-\mathbf{\sigma }_{\overline{p}}\right) \cdot 
\mathbf{n}\left\vert \psi _{1}\right\rangle =-\sqrt{2}e^{i\varphi }\sin
\theta \left\vert \psi _{P}\right\rangle ,  \label{c1}
\end{equation}%
\begin{equation}
\left( \mathbf{\sigma }_{p}-\mathbf{\sigma }_{\overline{p}}\right) \cdot 
\mathbf{n}\left\vert \psi _{0}\right\rangle =2\cos \theta \left\vert \psi
_{P}\right\rangle ,  \label{c2}
\end{equation}%
\begin{equation}
\left( \mathbf{\sigma }_{p}-\mathbf{\sigma }_{\overline{p}}\right) \cdot 
\mathbf{n}\left\vert \psi _{-1}\right\rangle =\sqrt{2}e^{-i\varphi }\sin
\theta \left\vert \psi _{P}\right\rangle ,  \label{c3}
\end{equation}%
\begin{eqnarray}
&&\left( \mathbf{\sigma }_{p}-\mathbf{\sigma }_{\overline{p}}\right) \cdot 
\mathbf{n}\left\vert \psi _{P}\right\rangle  \label{c4} \\
&=&2\cos \theta \left\vert \psi _{0}\right\rangle -\sqrt{2}e^{-i\varphi
}\sin \theta \left\vert \psi _{1}\right\rangle +\sqrt{2}e^{i\varphi }\sin
\theta \left\vert \psi _{-1}\right\rangle .  \notag
\end{eqnarray}%
It is noticeable that an ortho-state in a brane convert into a para-state
only in the second brane and vice versa. Then, considering only the
ortho-state in our brane \footnote{%
As shown in the following, the branching ratio of invisible decays of
ortho-states is larger by many orders of magnitude by contrast to this of
para-states.}, the relevant wave function to consider for matter exchange
can be written as:%
\begin{equation}
\left\vert \Psi \right\rangle =\left( 
\begin{array}{c}
\left\vert \Psi _{+}\right\rangle \\ 
\left\vert \Psi _{-}\right\rangle%
\end{array}%
\right) =\left( 
\begin{array}{c}
\left\vert \Psi _{O}\right\rangle \\ 
\left\vert \Psi _{P}\right\rangle%
\end{array}%
\right) =\left( 
\begin{array}{c}
\left\vert \psi _{1}\right\rangle \\ 
\left\vert \psi _{0}\right\rangle \\ 
\left\vert \psi _{-1}\right\rangle \\ 
\left\vert \psi _{P}\right\rangle%
\end{array}%
\right) ,  \label{Wfunc}
\end{equation}%
where $\left\vert \Psi _{\pm }\right\rangle $ are the wave functions in each
brane $(+)$ or $(-)$, and $\left\vert \Psi _{O/P}\right\rangle $ the wave
functions for the ortho-state and para-state. The two-brane wave function $%
\left\vert \Psi (t)\right\rangle $ then follows: 
\begin{equation}
i\hbar \partial _{t}\left\vert \Psi (t)\right\rangle =(\mathbf{H}_{0}+%
\mathbf{W)}\left\vert \Psi (t)\right\rangle ,  \label{Sch}
\end{equation}%
where the Hamiltonian $\mathbf{W}$ can be recast as:%
\begin{equation}
\mathbf{W}=\hslash \Omega \left( 
\begin{array}{cc}
\mathbf{0} & \left\vert \mathbf{C}\right\rangle \\ 
\left\langle \mathbf{C}\right\vert & 0%
\end{array}%
\right) \text{\ where }\left\vert \mathbf{C}\right\rangle =\left( 
\begin{array}{c}
i\sqrt{2}e^{i\varphi }\sin \theta \\ 
-2i\cos \theta \\ 
-i\sqrt{2}e^{-i\varphi }\sin \theta%
\end{array}%
\right) ,  \label{newW}
\end{equation}%
and where $\mathbf{H}_{0}=$ diag$(E_{O}-i\hslash \Gamma _{O},E_{O}-i\hslash
\Gamma _{O},E_{O}-i\hslash \Gamma _{O},E_{P}-i\hslash \Gamma _{P})$. $E_{O}$
(respectively $E_{P})$ is the eigenenergy and $\Gamma _{O}$ (respectively $%
\Gamma _{P}$) is the decay rate of the ortho-state (respectively of the
para-state).

$\left\vert \Psi (t)\right\rangle $ is obtained from the Lippmann--Schwinger
equation:%
\begin{equation}
\left\vert \Psi (t)\right\rangle =\left\vert \Psi ^{(0)}(t)\right\rangle
+\int_{-\infty }^{+\infty }G^{(0)}(t-t^{\prime })\mathbf{W}\left\vert \Psi
(t^{\prime })\right\rangle dt^{\prime },  \label{LPE}
\end{equation}%
where the propagator follows $\left( i\hbar \partial _{t}-\mathbf{H}%
_{0}\right) G^{(0)}(t-t^{\prime })=\mathbf{1}\delta (t-t^{\prime })$ such
that $G^{(0)}(t-t^{\prime })=-i\hbar ^{-1}e^{-i\hbar ^{-1}\mathbf{H}%
_{0}(t-t^{\prime })}\Theta (t-t^{\prime })$,$\ $with $\Theta (t)$ the
Heaviside step function. $\left\vert \Psi ^{(0)}(t)\right\rangle =i\hbar
G^{(0)}(t)\left\vert \Psi ^{(0)}\right\rangle $ is the solution of Eq. \ref%
{Sch} when $\mathbf{W}=0$ with%
\begin{equation}
\left\vert \Psi ^{(0)}\right\rangle =\left( 
\begin{array}{c}
\left\vert \Psi _{O}^{(0)}\right\rangle \\ 
0%
\end{array}%
\right) \text{ where }\left\vert \Psi _{O}^{(0)}\right\rangle =\left( 
\begin{array}{c}
p \\ 
q \\ 
r%
\end{array}%
\right) \text{,}  \label{P0}
\end{equation}%
with $\left\vert p\right\vert ^{2}+\left\vert q\right\vert ^{2}+\left\vert
r\right\vert ^{2}=1$. $\left\vert \Psi ^{(0)}\right\rangle $ defines the
initial state when the particle is created. Assuming that $\hbar \Omega \ll
E_{O},E_{P}$ we use an up to the second-order expansion of the
Lippmann--Schwinger equation such that: 
\begin{equation}
\left\vert \Psi (t)\right\rangle \sim \left\vert \Psi ^{(0)}(t)\right\rangle
\label{Serie}
\end{equation}%
\begin{equation*}
+\int_{-\infty }^{+\infty }G^{(0)}(t-t^{\prime })\mathbf{W}\left\vert \Psi
^{(0)}(t^{\prime })\right\rangle dt^{\prime }
\end{equation*}%
\begin{equation}
+\int_{-\infty }^{+\infty }\int_{-\infty }^{+\infty }G^{(0)}(t-t^{\prime })%
\mathbf{W}G^{(0)}(t^{\prime }-t^{\prime \prime })\mathbf{W}\left\vert \Psi
^{(0)}(t^{\prime \prime })\right\rangle dt^{\prime \prime }dt^{\prime }. 
\notag
\end{equation}%
\newline

We are now going to calculate the whole decay rate $\Gamma $ of the $%
\overline{p}p$ bound state in the two-brane Universe. We set $\Gamma
=(1/2)\tau ^{-1}$ where $\tau $ is the mean lifetime of the $\overline{p}p$
state given by $\tau =\int_{0}^{\infty }tf(t)dt$ where the related
distribution function for the probability to observe the particle is $%
f(t)=-(d/dt)P$ with $P=\left\langle \Psi (t)\right. \left\vert \Psi
(t)\right\rangle $. Then, we get: $\Gamma ^{-1}=2\int_{0}^{\infty
}\left\langle \Psi (t)\right. \left\vert \Psi (t)\right\rangle dt$. Using
Eq. \ref{Serie}, we can now express $\left\langle \Psi (t)\right. \left\vert
\Psi (t)\right\rangle $ up to the second order of approximation and we can
simply compute $\Gamma ^{-1}$: 
\begin{eqnarray}
\Gamma ^{-1} &\sim &\frac{1}{\Gamma _{O}}+\frac{\Gamma _{O}^{2}-\Gamma
_{P}^{2}}{\Gamma _{P}\Gamma _{O}^{2}}\left\vert \left\langle \mathbf{C}%
\right. \left\vert \Psi _{O}^{(0)}\right\rangle \right\vert ^{2}
\label{Gamma-1} \\
&&\times \frac{\Omega ^{2}}{\hbar ^{-2}\left( E_{O}-E_{P}\right)
^{2}+(\Gamma _{O}+\Gamma _{P})^{2}},  \notag
\end{eqnarray}%
where $\left\vert \left\langle \mathbf{C}\right. \left\vert \Psi
_{O}^{(0)}\right\rangle \right\vert ^{2}=\left\vert \sqrt{2}\left(
e^{i\varphi }r-e^{-i\varphi }p\right) \sin \theta +2q\cos \theta \right\vert
^{2}$.

It is noticeable that the result depends on the polarization state of the
ortho-state compared with the direction $\mathbf{n}$ of the effective
magnetic field $\mathbf{B}_{\bot }=\mathbf{n}B_{\bot }$ while it is not the
case for a neutron for instance \cite{pheno} \footnote{%
For a neutron with a given polarization state such that $\left\vert
u\right\rangle =\alpha \left\vert \uparrow \right\rangle +\beta \left\vert
\downarrow \right\rangle $ (with $\left\vert \alpha \right\vert
^{2}+\left\vert \beta \right\vert ^{2}=1$), considering the off-diagonal
terms of $\mathbf{H}_{c}$ proportional to $\mathbf{\sigma }\cdot \mathbf{n}$%
, we get $\mathbf{\sigma }\cdot \mathbf{n}\left\vert u\right\rangle =\alpha
^{\prime }\left\vert \uparrow \right\rangle +\beta ^{\prime }\left\vert
\downarrow \right\rangle $. Although $\left( \alpha ^{\prime },\beta
^{\prime }\right) \neq \left( \alpha ,\beta \right) $ in most cases (i.e.
the polarization is modified when the neutron leaps into the hidden brane),
one gets: $\left\vert \alpha ^{\prime }\right\vert ^{2}+\left\vert \beta
^{\prime }\right\vert ^{2}=1$ such that the swapping amplitude does not
depend on the polarization state of the neutron. It is not longer the case
of the $p\overline{p}$ states here under consideration due to the structure
of their wave functions.}. Anyway, in most experiments, there is no way for
the particles to be produced with a fixed polarization. In this case, $%
\left\vert \left\langle \mathbf{C}\right. \left\vert \Psi
_{O}^{(0)}\right\rangle \right\vert ^{2}=4/3$ for unpolarized particles.
From Eq. \ref{Gamma-1} we can deduce the whole decay rate $\Gamma $:%
\begin{equation}
\Gamma =\Gamma \left( \text{o-}p\overline{p}\rightarrow \text{visible decay}%
\right) +\Gamma \left( \text{o-}p\overline{p}\rightarrow \text{invisible}%
\right) ,  \label{decayR}
\end{equation}%
with 
\begin{equation}
\Gamma \left( \text{o-}p\overline{p}\rightarrow \text{visible decay}\right)
=\Gamma _{O}\left( 1-\frac{\Gamma _{O}}{\Gamma _{P}}\mathcal{R}\right) ,
\label{G10}
\end{equation}%
and%
\begin{equation}
\Gamma \left( \text{o-}p\overline{p}\rightarrow \text{invisible}\right)
=\Gamma _{P}\mathcal{R},  \label{G20}
\end{equation}%
and where%
\begin{equation}
\mathcal{R}=\left\vert \left\langle \mathbf{C}\right. \left\vert \Psi
_{O}^{(0)}\right\rangle \right\vert ^{2}\frac{\Omega ^{2}}{\hbar ^{-2}\left(
E_{O}-E_{P}\right) ^{2}+(\Gamma _{O}+\Gamma _{P})^{2}}.  \label{probaR}
\end{equation}%
Thus, the invisible decay rate (see Eq. \ref{G20}) is simply the para-state
decay rate (since o-$p\overline{p}$ converts into p-$p\overline{p}$ in the
hidden brane) weighted by $\mathcal{R}$, the probability of swapping from our visible
braneworld to the hidden brane. By contrast, Eq. \ref{G10} shows that o-$p%
\overline{p}$ decay products -- recorded in our visible world -- must occur at a
lower rate since some particles are lost through decays in the hidden brane. At last,
Eq. \ref{G20} shows that some decays must occur as invisible
ones leading to an apparent loss of energy which is more easily detectable
than a change of the visible o-$p\overline{p}$ decay rate.

Finally, we simply deduce the branching ratio for the invisible decay of
unpolarized particles: 
\begin{eqnarray}
\text{Br}(\text{o-}p\overline{p} &\rightarrow &\text{invisible})=\frac{4}{3}%
\frac{\Gamma _{P}}{\Gamma _{O}}  \label{Br} \\
&&\times \frac{\Omega ^{2}}{\hbar ^{-2}\left( E_{O}-E_{P}\right)
^{2}+(\Gamma _{O}+\Gamma _{P})^{2}}  \notag
\end{eqnarray}%
As a remark, similar calculations can be done in order to obtain the
branching ratio Br$($p-$p\overline{p}\rightarrow $invisible$)$ for the
invisible decay of para-states. As a result, one then gets:%
\begin{equation}
\frac{\text{Br}(\text{p-}p\overline{p}\rightarrow \text{invisible})}{\text{Br%
}(\text{o-}p\overline{p}\rightarrow \text{invisible})}=3\left( \frac{\Gamma
_{O}}{\Gamma _{P}}\right) ^{2}\text{,}  \label{compare}
\end{equation}%
such that Br$($p-$p\overline{p} \rightarrow $invisible$)$ is up to $6$
(respectively $5$ and $4$) orders of magnitude smaller than Br$($o-$p%
\overline{p}\rightarrow $invisible$)$ for positronium (respectively
charmonium and bottonium), thus justifying to focus only on invisible decays
of ortho-states.

We are now investigating the ortho-positronium (o-Ps) (respectively the $%
J/\psi $ charmonium and the $\Upsilon $ bottonium) invisible decay through
its conversion into para-positronium (p-Ps) (respectively $\eta _{c}$
charmonium and $\eta _{b}$ bottonium) in the hidden brane.

Equation \ref{Br} applies by using the relevant parameters (see table \ref%
{t1}). Here the difference $E_{O}-E_{P} $ includes the following
contributions: 
\begin{equation}
E_{O}-E_{P}=\left( m_{O}-m_{P}\right) c^{2}+m_{O}V_{+}-m_{P}V_{-}
\label{Diff}
\end{equation}%
where $m_{O}$ and $m_{P}$ are the masses of the ortho-state and of the
para-state respectively, and $V_{\pm }$ are the gravitational potentials in
each brane. Doing this, we consider the following hypotheses:\newline

(i) \textit{Null electric dipole moment hypothesis}

In our previous works \cite{coupl,npmth,exp}, it has been underlined that
the interaction between the magnetic vector potential and the charge of a
particle precludes the particle swapping between branes. As a consequence,
since we consider electron/positon pairs or quark/antiquark pairs, we must
consider these pairs as strictly chargeless and without charge structure.
Let $E_{bind}$ be the typical binding energy of the fermion/antifermion
pair. Then, the time $\Delta t\sim \hbar /E_{bind}$ during which quantum
fluctuations allow for an instantaneous electric dipole moment (EDM) must be
smaller than the period $T\sim \hbar /\left( E_{O}-E_{P}\right) $ of the
Rabi oscillation during which the $p\overline{p}$ state oscillates between
each brane. This allows a time averaged EDM equal to zero. That means that $%
E_{O}-E_{P}\ll E_{bind}$ must be verified. This condition is verified in
quarkonium where the mass-energy difference between the ortho and para
states dominates (i.e. $E_{O}-E_{P}\sim \left( m_{O}-m_{P}\right) c^{2}$).
Indeed, $E_{bind}\approx 1$ GeV and $\approx 0.7$ GeV while $%
E_{O}-E_{P}\approx 61$ MeV and $\approx 114$ MeV for the bottonium and the
charmonium respectively \cite{Qk}. By contrast, for the positronium the
gravitational potential is now dominant (i.e. $E_{O}-E_{P}\sim m_{Ps}\left(
V_{+}-V_{-}\right) $) and $E_{bind}\approx 6.8$ eV \cite{Psqu}. Then,
positronium is a relevant probe if $\left\vert V_{+}-V_{-}\right\vert
\lesssim 10^{-6}c^{2}$. The consequence of this is discussed later in this
paper.\newline

\begin{table}[ht]
\centering
\renewcommand{\arraystretch}{1.5} 
\begin{tabular}{l|c||l|c}
$p\overline{p}$ & $\hbar \Omega $ & p (or $\overline{p})$ & $g$ \\ 
\hline\hline
o-Ps & $<6.4\times 10^{-4}$ eV & $e^{-}/e^{+} $ & $<1.1\times 10^{-8}$
m$^{-1}$ \\ 
$J/\psi$ & $<0.15$ MeV & $c/\overline{c}$ & $<1.2\times 10^{4}$ m$^{-1}$
\\ 
$\Upsilon$ & $<0.10$ MeV & $b/\overline{b}$ & $<4.8\times 10^{4}$ m$^{-1}$ \\ 
\hline\hline
\textit{n} (udd) & $<0.14$ eV & u or d & $<2.4\times 10^{-3}$ m$^{-1}$ \\ \hline
\end{tabular} 
\caption{Derived constraints on $\hbar \Omega $ for positronium and
quarkonium, and on the interbrane coupling constant $g$ for some Standard
Model particles. The quark constituent masses used are $m_{b}=4730$ MeV and $%
m_{c}=1550$ MeV \cite{CQM1,CQM2,CQM3,CQM4,Qk}. Neutron results are
derived from previous experimental results \cite{npm} and given for comparison.}
\label{t2}
\end{table}

(ii) \textit{Global gravitational potential hypothesis}

The gravitational potential $V_{\pm }$ felt by the particles incorporates
many contributions. Some are local and others are more global at a
cosmological scale. Matter distribution can be considered as homogeneous and
isotropic at a cosmological scale. If the second brane is similar to our
own, then the difference of the gravitational contributions of each brane
vanishes when considering large scale contributions only. As a consequence,
the difference of the gravitational contributions of each brane just depends
on fluctuations of local mass distributions in each brane at various scales.
The first contribution comes from matter fluctuations related to large-scale
galactic superstructures \cite{s8,rels8}. The amplitude of such fluctuations
is characterized by the well-known parameter $\sigma _{8}$ \cite{s8,rels8},
which allows to estimate the amplitude $V_{C}$ of the gravitational
potential at the level of large-scale superstructures \cite{rels8}. Here we
get \cite{rels8}: $V_{C}=(3/2)\Omega _{m0}\left( 8\text{ Mpc}%
/H_{0}^{-1}\right) ^{2}$ $\sigma _{8}\approx 10^{-5}c^{2}$ (where $\Omega
_{m0}$ is the present-day density parameter for baryons and cold dark
matter, and $H_{0}$ the Hubble constant). At a closer scale, Milky Way
contributes for $V_{MW}\approx 5\times 10^{-7}c^{2}$, while the Sun, the
Earth, and the Moon provide lower contributions of about $10^{-8}c^{2}$, $%
7\times 10^{-10}c^{2}$, and $10^{-13}c^{2}$, respectively \cite{pheno}.
Considering hidden branes similar to our own, or with a lower matter
content, $V_{C\text{ }}$ should allow to define an upper constraint on $%
|V_{+}-V_{-}| < 10^{-5}c^{2}$.\newline

(iii) \textit{Global magnetic vector potential hypothesis}

The overall ambient astrophysical magnetic vector potential $\mathbf{A}%
_{amb} $ was previously discussed in the literature in a different context 
\cite{vp1,vp2,vp3,vp4,vp5}. $\mathbf{A}_{amb}$ is the sum of all of the
magnetic vector potential contributions related to the magnetic fields of
astrophysical objects (planets, stars, galaxies, etc.) since $\mathbf{B}(%
\mathbf{r})=\mathbf{\nabla }\times \mathbf{A}(\mathbf{r})$. At large
distances from sources (for instance, close to the Earth), $\mathbf{A}_{amb}$
is almost uniform (i.e. $\mathbf{\nabla }\times \mathbf{A}_{amb}\approx 
\mathbf{0}$) and cannot be cancelled out with magnetic shields \cite{vp5}.
As a rule of thumb, $A\approx DB$ where $D$ is the distance from the
astrophysical source and $B$ is the typical field induced by the object. The
expected order of magnitude of $A_{amb}$ can be then roughly constrained.
Galactic magnetic field variations on local scales towards the Milky Way
core ($A_{amb}\approx 2\times 10^{9}$ T m) are usually assumed \cite%
{vp1,vp2,vp3,vp4,vp5}. By contrast, the Earth's magnetic field leads to $200$
T m while the Sun contributes $10$ T m \cite{vp5}. By contrast, the
magnitude of intergalactic contributions are less clear \cite{pheno}). Then,
for now $A_{amb}$ can be fairly bounded by the lower estimated value: $%
A_{amb} = 10^{9}$ T m.\newline

\begin{figure}[ht]
\centering
\includegraphics[width=8.5 cm]{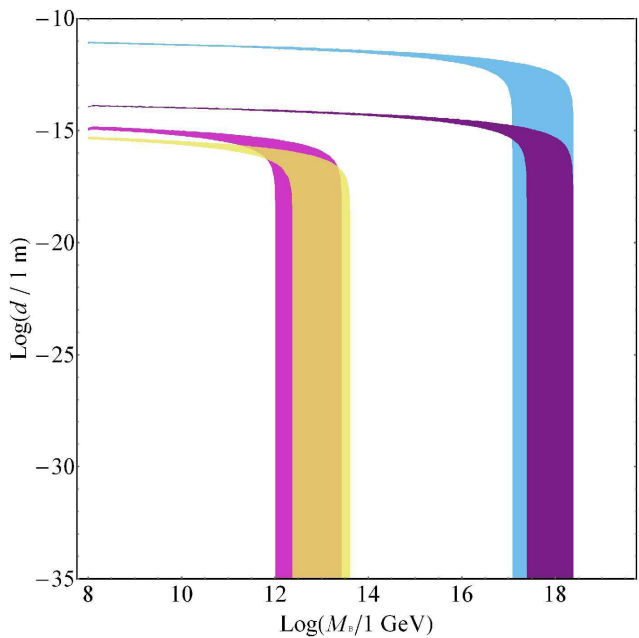}
\caption{(Color online). Bounds on the brane energy scale $M_{B}$ and on the
interbrane distance $d$ derived from each particle kind: bottonium (yellow),
charmonium (magenta), positronium (cyan) and neutron (purple). Each domain
on the left of a colored band is excluded. Each colored band widens to the
right when sensitivity increases: up to the Planck scale for neutron and
positronium ($\text{Br}(\text{o-Ps} \rightarrow \text{invisible}) = 1.4
\times 10^{-6}$), up to $\text{Br}(\text{quarkonium} \rightarrow \text{%
invisible}) = 10^{-6}$ for quarkonia.}
\end{figure}

As a result, from table \ref{t1} and Eq. \ref{Br} we deduce the results
shown in table \ref{t2}. Table \ref{t2} gives the upper bounds for the
coupling constants ($\hbar \Omega $ or $g$) both for bound states and some
fundamental particles of the Standard Model. From table \ref{t2} data and
Eq. \ref{cg}, we can now derive lower bounds on the brane energy scale $%
M_{B} $ and upper bounds on the interbrane distance $d$ -- for a $SO(3,1)$%
-broken $M_{4}\times R_{1}$bulk -- which are independent of the particle
under consideration. The results are summarized on Fig. 1. Each domain on
the left of a colored band is excluded: If matter disappearance can occur in
the present braneworld scenario, these values of $(M_{B},d)$ are not
relevant. One notes that bounds are much poorer for quarkonia than for
positronium or neutron. For the latter, it is noticeable that the
constraints are just one order of magnitude below the reduced Planck energy
scale, except if one accepts as possible a fine tuning of the distance $d$
between $1$ fm or $1$ \AA\ allowing for significantly low values for $M_{B}$
(see Fig. 1). To be more quantitative, table \ref{t3} (left column) gives
the bounds of the brane energy scale $M_{B}$ for each particle.

\begin{table}[ht]
\centering
\renewcommand{\arraystretch}{1.5} 
\begin{tabular}{l|c||c}
Particle & $M_{B}$ & $\text{Br}(\text{o-}p \overline{p}\rightarrow \text{%
invisible})$ \\ \hline\hline
o-Ps & $>1.2\times 10^{17}$ GeV & $1.4\times 10^{-6}$ \\ 
$J/\psi $ & $>1.0\times 10^{12}$ GeV & $1.3\times10^{-16}$ \\ 
$\Upsilon $ & $>2.4\times 10^{12}$ GeV & $2.9\times 10^{-16}$ \\ \hline
neutron & $>2.5\times 10^{17}$ GeV &  \\ \hline
\end{tabular}
\caption{Left column: Various bounds on the brane energy scale $M_{B}$
derived from each particle kind for interbrane distances below $0.5$ fm.
Right column: Expected branching ratios for a brane energy scale at the
reduced Planck energy ($2.43\times10^{18}$ GeV).} 
\label{t3}
\end{table}

Considering the same two-brane scenario and assuming now that the brane
energy scale is the reduced Planck energy scale, we can derive the expected
branching ratio for the positronium and quarkonia invisible decays (see
right column in table \ref{t3}), which need to be reached to observe the
phenomenon. While the expected branching ratio for quarkonia is far beyond
any current experimental skill, it is not the case for positronium which
could compete with neutron to constrain braneworld scenarios \cite%
{npm,npmth,exp}. In addition, in Fig. 1, each colored band widens to the
right when sensitivity increases: up to the Planck scale for neutron and
positronium ($\text{Br}(\text{o-Ps} \rightarrow \text{invisible}) = 1.4
\times 10^{-6}$), up to $\text{Br}(\text{quarkonium} \rightarrow \text{%
invisible}) = 10^{-6}$ for quarkonia.

As a major result, pretty low-energy experiments -- involving positronium or
neutron -- appear then more suitable to probe the Planck scale than
experiments needing colliders to produce quarkonia states. Even for a brane
energy scale $M_{B}$ at the Planck energy, low-energy disappearance
phenomena can occur and could be observed in future low-energy experiments
while the Planck scale is directly unreachable with colliders. However, to
occur, positronium oscillations imply that the gravitational potentials in
each brane must verify $\left\vert V_{+}-V_{-}\right\vert \lesssim
10^{-6}c^{2}$, i.e. a value one order of magnitude lower that the upper
gravitational constraint mentioned above. As a result, positronium and
neutron experiments are complementary as they could allow to discriminate
between two kinds of gravitational environment.

\section*{Acknowledgements}

C.S. is supported by a FRIA doctoral grant from the Belgian F.R.S-FNRS. The
authors thank Nicolas Reckinger for reading the manuscript.

\end{document}